\documentclass[]{piparticle-final}
\usepackage{graphicx}
\usepackage{amsmath}
\usepackage{epstopdf}
\newcommand{\figA}{
\begin{figure}
\begin{center}
\includegraphics[width=0.48\textwidth]{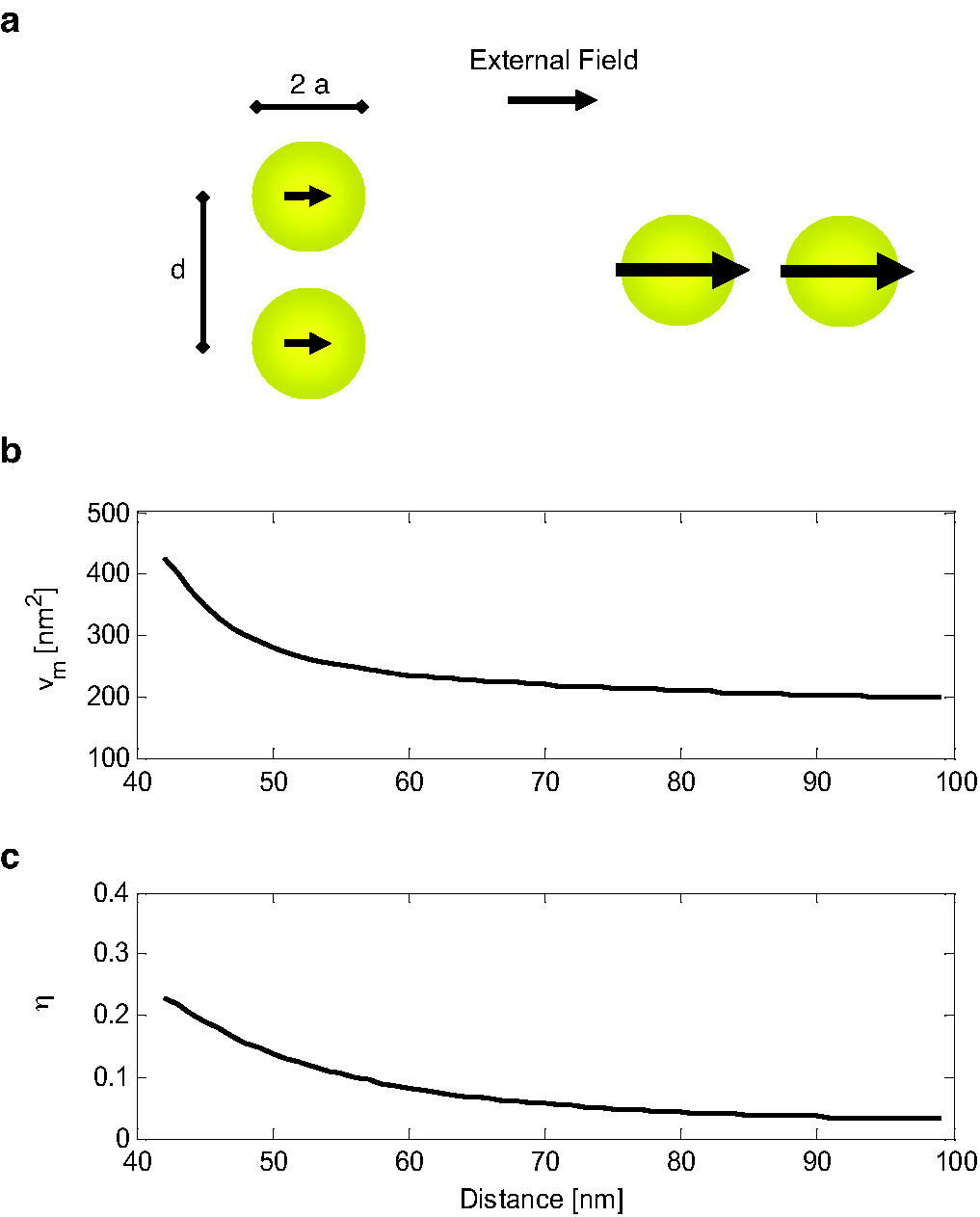}
\end{center}
\caption{
Conceptual idea of the technique. (a) Two metallic
nanoparticles of radii $a$ located at a distance $d$ are illuminated with a linearly
polarized electromagnetic field. (b) Theoretical results as a function of the
interparticle distance for the average $C_{sca}$ over all polarizations ($v_m$)
and (c) the anisotropy. Parameters of the calculation: $a$ = 20 nm, wavelength of the light: 532 nm.
} \label{figure1}
\end{figure}
}

\newcommand{\figB}{
\begin{figure}
\begin{center}
\includegraphics[width=0.48\textwidth]{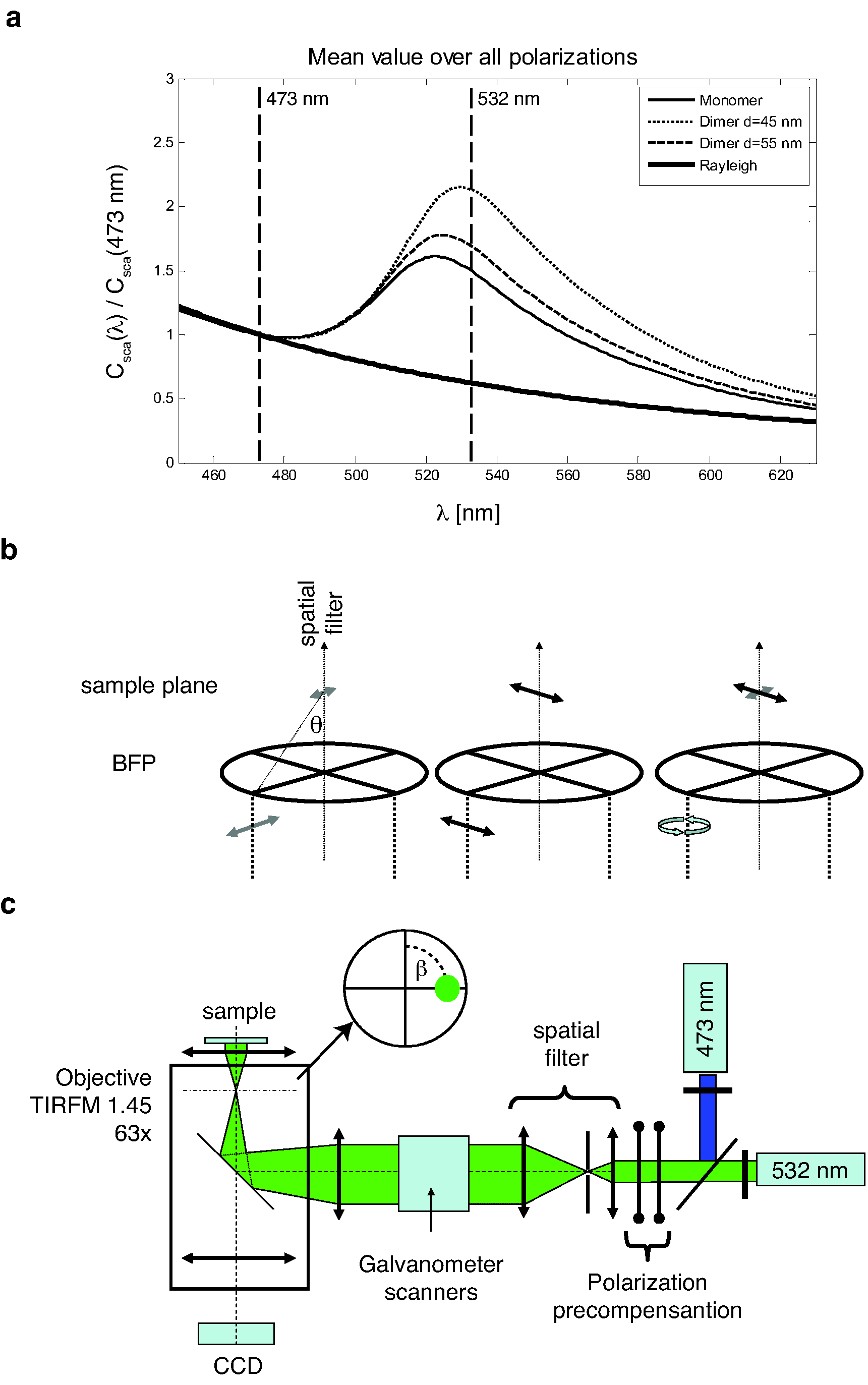}
\end{center}
\caption{
Experimental Setup.
(a) Comparison of spectra averaging over all polarizations. While the spectrum of dielectric
particles (thick solid line) decreases monotonically, the spectra of metallic monomers and dimers (thin lines) show a plasmon resonance.
(b) The polarization of the refracted wave is dependent on the direction of the incident polarization with respect to the displacement in the back focal plane (BFP) of the objective which defines the plane of incidence.
(c) The sample is illuminated in Total Internal Reflection and imaged using a cooled CCD. Laser light is tightly focused off-center in the back
focal plane of the objective and the angle $\beta$ is moved using a pair of galvanometer scanners moving in orthogonal directions.
} \label{figure2}
\end{figure}
}

\newcommand{\figC}{
\begin{figure}
\begin{center}
\includegraphics[width=0.48\textwidth]{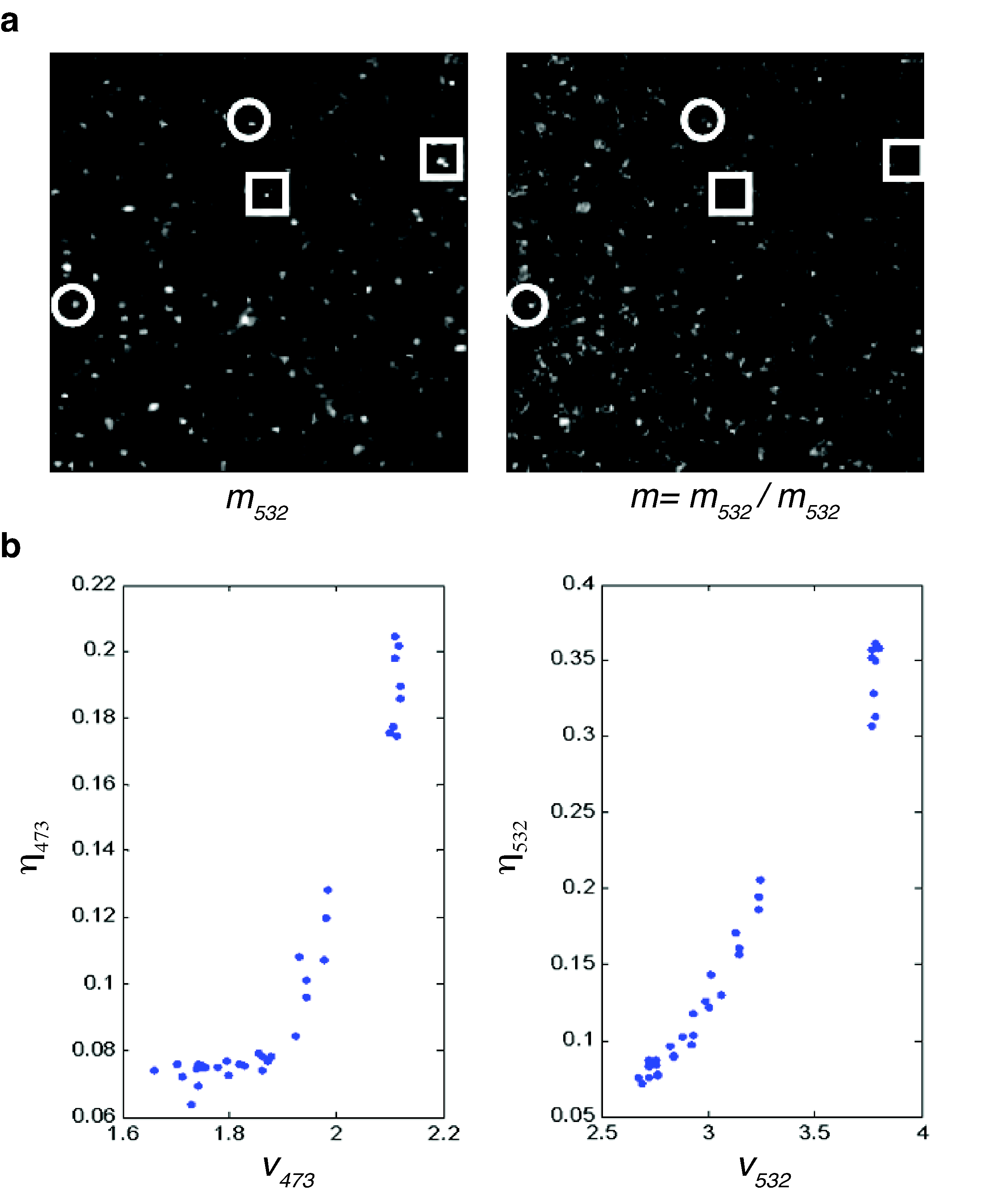}
\end{center}
\caption{
Experimental results 
(a) Representative images of a field of view for single wavelength (left) and a ratio (right) imaging. 
Some points (circles) are bright in both images (gold) while others squares faint in the ratio image (not gold).
Gold containing regions were segmented by finding bright pixels in both images.
Images size: 50x50 ${\mu m}^2$. 440x440 pixels.
(b) Anisotropy vs. mean value. 
A strong correlation is observed between anisotropy and mean value for both wavelengths as expected. 
The 473 nm data shows a plateau due to the intrinsic anisotropy of the system.
} \label{figure3}
\end{figure}
}

\newcommand{\figD}{
\begin{figure}
\begin{center}
\includegraphics[width=0.48\textwidth]{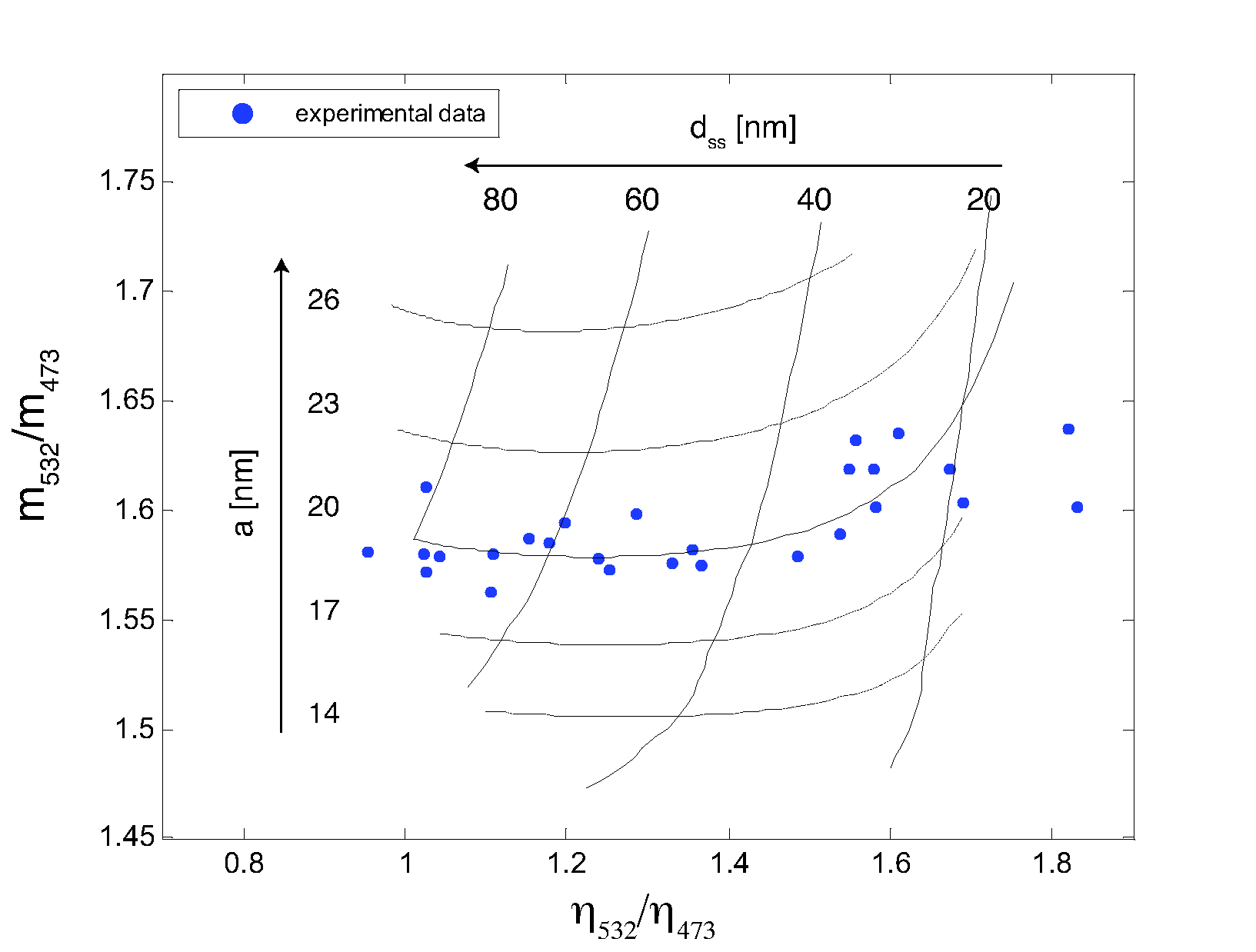}
\end{center}
\caption{
Comparison between experimental data and homodimer model.
The mean value ratio is plotted against the anisotropy ratio for the regions segmented from the images (blue dots). Theoretical calculations for different homodimers are also plotted.
The vertical lines show the results keeping constant the surface to surface distance ($d_{ss}$) while changing the radii ($a$). 
The opposite is shown in the horizontal lines.
Remarkably, experimental data distributed close to the curve for 20 nm homodimers (solid line) as expected as this is in fact the mean radii of the particles used.
Notice that this is not a fit (no free parameters), but the predictions from the homodimer model superimposed to the experimental data. 
} \label{figure4}
\end{figure}
}

\newcommand{\figE}{
\begin{figure}
\begin{center}
\includegraphics[width=0.48\textwidth]{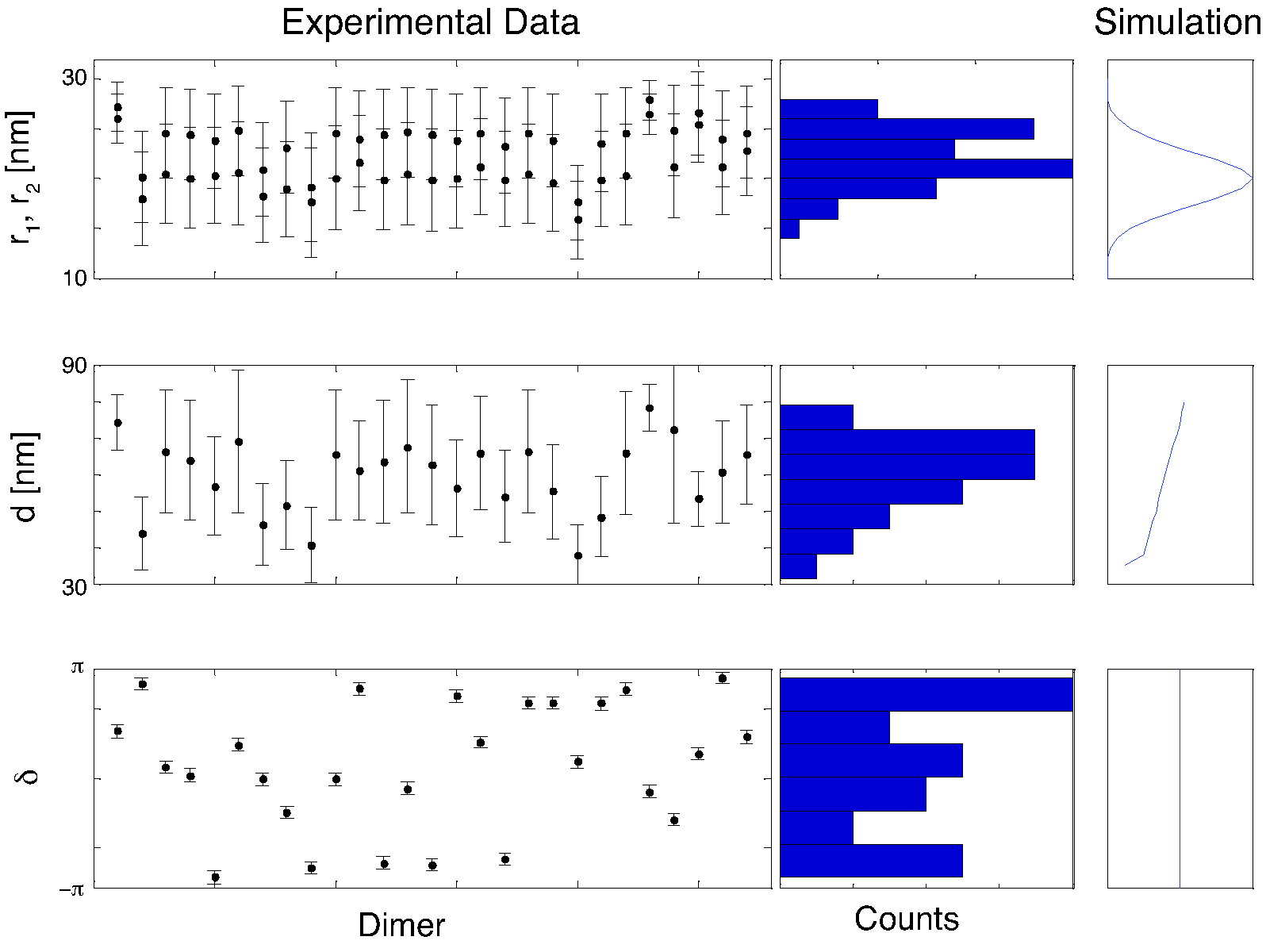}
\end{center}
\caption{
Fit to a heterodimer configuration using GMMie.
For each dimer (left column, x axis), the radii (top), the distance (middle) and the angle (bottom) were obtained.
Experimental and simulated histograms are shown for each magnitude (middle and right columns).
} \label{figure5}
\end{figure}
}

\usepackage{cite}

\begin{document}
\volume{2}               
\articlenumber{020010}   
\journalyear{2010}       
\editor{V. Lakshminarayanan}   
\reviewers{S. Roy, Dayalbagh Educational Institute, Agra, India.}  
\received{$22$ April 2010}     
\accepted{2 December 2010}   
\runningauthor{H. E. Grecco \itshape{et al.}}  
\doi{020010}         

\title{Experimental determination of distance and orientation of metallic nanodimers by polarization
dependent plasmon coupling}

\author{H. E. Grecco,\cite{inst1, inst2}\thanks{E-mail: hgrecco@df.uba.ar} \hspace{0.5em}O. E. Mart\'{i}nez\cite{inst1}\thanks{E-mail: oem@df.uba.ar}}

\pipabstract{ 
Live cell imaging using metallic nanoparticles as tags is an emerging technique to visualize long and highly dynamic processes due to the lack of photobleaching and high photon rate.
However, the lack of excited states as compared to fluorescent dyes prevents the use of resonance energy transfer and recently developed super resolution methods to measure distances between objects closer than the diffraction limit.
In this work, we experimentally demonstrate a technique to determine subdiffraction distances based on the near field coupling of metallic nanoparticles.
Due to the symmetry breaking in the scattering cross section, not only distances but also relative orientations can be measured. Single gold nanoparticles were prepared on glass, statistically yielding a small fraction of dimers.
The sample was sequentially illuminated with two wavelengths to separate background from nanoparticle scattering based on their spectral properties.
A novel total internal reflection illumination scheme in which the polarization can be rotated was used to further minimize background contributions.
In this way, radii, distance and orientation were measured for each individual dimer, and their statistical distributions were found to be in agreement with the expected ones.
We envision that this technique will allow fast and long term tracking of relative distance and orientation in biological processes.
}

\maketitle

\blfootnote{
\begin{theaffiliation}{99}
   \institution{inst1} Laboratorio de Electr\'{o}nica Cu\'{a}ntica, Universidad de Buenos Aires. Buenos Aires, Argentina.
   \institution{inst2} Current address:
 Department of Systemic Cell Biology Max Planck Institute of Molecular Physiology.
Dortmund, Germany \end{theaffiliation}
}

\section{Introduction}
Microscopy is an example of the ongoing symbiotic relationship between physics and biology: as early microscopes allowed fundamental discoveries like microorganisms or DNA; the need to see smaller, faster and deeper has pushed the development of a plethora of optical concepts and microscopy techniques.
Today, fluorescence microscopy is an essential tool in biology as it can visualize the spatio-temporal dynamics of intracellular processes.
However, many important mechanisms, like protein interaction, clustering or conformational changes, occur at length scales smaller than the resolution limit of conventional microscopy and therefore cannot be assessed by standard imaging.
Unraveling the dynamics of such inter- and intramolecular mechanisms that provides function richness to molecules and molecular complexes is essential to understand key biological processes such as cellular signal propagation.

Subdiffraction distances have been determined by exploiting quantum and near field properties of the interaction between light and matter in the nanometer scale.
For example, Fluorescence/F\"orster Resonance Energy Transfer (FRET) \cite{haj_imaging_2002,jares-erijman_fret_2003,kam_mapping_1995,mahajan_bcl-2_1998} has proven to be a valuable technique as it provides an optical signal directly related to the proximity of the molecules.
The desire to extend this technique to other biological systems with different time and length scales has been hindered by the inherent limitations of fluorescent dyes (i.e. lack of photostability, low brightness and short range of interaction). 
Super resolution techniques such as STED \cite{VolkerWestphal04112008} or PALM \cite{EricBetzig09152006} have recently gained momentum to directly observe fluorescent molecules spaced closer than the diffraction limit.
Although much work has been done to increase the total acquisition time and frame rate, these methods are still limited by the lack of photostability and the need to image a single resolvable structure per diffraction limited spot at a time.

In the past, it has been shown that scattering microscopy using metallic nanoparticles can complement its fluorescence sibling as it uses an everlasting tag with no rate-limited amount of photons \cite{lasne_single_2006}. 
Metallic nanoparticles are stable, biocompatible and easy to synthesize and conjugate to biological targets and thereby ideal as contrast agents.
A landmark example of the biological application of such techniques was the direct observation of receptors hopping across previously unknown membrane domains. This provided valuable insight into the spatial regulation of signaling complexes and closed a 30-year controversy about the diffusion coefficients of membrane proteins \cite{suzuki_rapid_2005}.
While previous experiments using fluorescent tags yielded a diffusion coefficient in biological membranes much slower than the one observed in synthetic membranes, the fast acquisition speed (40 10$^3$ frames per second) enabled by scattering microscopy  showed that this is the result of a fast diffusion and a slow hopping rate between domains \cite{kusumi2005}.

In addition, the presence of a plasmon, i.e. a collective oscillation of the free electrons within the nanoparticle, converts metallic nanoparticles into very effective scatterers when illuminated at their resonance optical frequency. The resulting strong electromagnetic enhancement in the vicinity of the particle provides a near field effect that can be used to sense information about their surroundings such as the effective index of refraction or the presence of other scatterers \cite{kelly_optical_2003,mcfarland_single_2003}.
For example, it has been experimentally shown that the shift in the plasmon resonance can be used to determine the length of DNA molecules attached to a metallic nanoparticle \cite{liu_2006}.
Moreover, the coupling between two nanoparticles in close proximity produces an alteration of the plasmon spectra. This alteration has been used as a nanometric ruler to determine the distance between them \cite{snnichsen_molecular_2005,reinhard2005}.
In a previous work, \cite{grecco_distance_2006} we theoretically showed that the coupling between two nanoparticles is highly sensitive to the polarization of the external field. 
The scattering cross section ($C_{sca}$) is maximum when the incident polarization is parallel to the dimer orientation due to the reinforcement of the external field by the induced dipoles [Fig. 1(a)].
As the coupling decreases monotonically with the distance between nanoparticles, so does the average $C_{sca}$ over all polarizations ($v_m)$ [Fig. 1(b)] and the anisotropy [Fig. 1(c)] defined as:

\begin{align}\label{eq:anisdef}
\eta = \frac{C_{sca}^{\parallel} - C_{sca}^{\perp}}{C_{sca}^{\parallel}+C_{sca}^{\perp}}.
\end{align}
We have proposed, in our previous work, that by measuring the scattering cross section as a function of the incident polarization angle, the axis of the dimer and the distance between nanoparticles could be determined.
In this work, we provide experimental evidence supporting this concept by measuring gold nanodimers on a glass surface and we introduce a novel total internal reflection experimental setup that provides polarized illumination with a high NA objective.
\figA

\section{Materials and methods}

\subsection{Sample preparation}
Coverslips were cleaned by sonication at 50$^{\circ}$ for 20 minutes in Milli-Q water, and then sequentially immersed for 5 seconds in HFL 5\%, sodium bicarbonate and acetone (analytic level).
After the cleaning process, coverslips were dried and stored in a chamber overpressurized with nitrogen until further use.
Before sample preparation, a Parafilm chamber was assembled on top of the coverslip.
To create a hydrophilic surface, bovine serum albumina (BSA) in phosphate buffer solution (PBS) was incubated for 15 min and then rinsed with PBS.
Fluorescein-streptavidin in PBS (50 mg/ml) was then incubated for 30 min and rinsed with PBS, to obtain an adsorbed layer that was verified using confocal fluorescence microscopy.
Finally, a solution of biotinylated gold nanoparticles, nominal radius ($20\pm5$) nm (GB-01-40. EY Laboratories, USA), was incubated for 15 min and then rinsed by washing 5 times with PBS.
The concentration and incubation time where empirically chosen to provide a concentration about 1 nanoparticle/10 $\mu$m$^2$. As the particles are randomly distributed, it is expected to find many monomers, some dimers, and very few trimers and higher n-mers.
A negative control sample was prepared in the same way but omitting the incubation of gold nanoparticles.

\subsection{Dual color scheme}
Spurious reflections and scattering centers other than gold will produce unwanted bright spots in the images. Even thresholding the image taken at the resonance peak (532 nm) will result in many false positive regions.
The presence of a plasmon resonance in the scattering spectrum of gold nanoparticles was used as a signature to distinguish them.
The ratio between the scattering cross section at 532 nm and 473 nm was found to be larger than 1.4 for gold monomers [Fig. 2(a), solid thin line] using Mie theory \cite{bohren_absorption_1983} and even larger for dimers (dashed line) as calculated using \nolinebreak[4]{GMMie}, a multiparticle extension of the Mie theory \cite{xu_electromagnetic_1995,xu_electromagnetic_1997}.
In contrast, non metallic scattering centers lack of a plasmon resonance and therefore yield a smaller ratio between 532 nm and 473 nm $C_{sca}$ (solid thick line).
Therefore, by imaging at these two wavelengths and thresholding the ratio image above 1.4, the pixels containing gold nanoparticles were further segmented.

\figB

\subsection{Polarization control in Total Internal Reflection}
We used Total Internal Reflection (TIR) microscopy \cite{axelrod_total_1984} to restrict the illumination to the surface of the coverslip using an evanescent wave.
In objective-based TIR, the beam is focused off-axis in the back focal plane (BFP) of the objective to achieve  critical illumination.
The components of evanescent field are defined by the angle of incidence ($\theta$) and the incident polarization with respect to the plane of incidence.
Indeed, rotating the excitation polarization before entering the microscope does not produce a constant intensity in the sample plane as the transmission efficiency for the parallel polarization [Fig. 2(b), left] will be much smaller than for the perpendicular one [Fig. 2(b), center].
A circularly polarized beam before the objective results in an ``elliptically''\footnote{The electromagnetic field in the sample cannot be said to be strictly elliptically polarized as an evanescent field (not a propagating beam) is generated after the interface. Nevertheless, an elliptical rotation of the electric field is achieved.} polarized field which has the minor axis in the plane of incidence [Fig. 2(b), right].
The ratio between the major and minor axis of this evanescent ``elliptical'' beam depends on $\theta$ and if the plane of incidence is changed, the ellipse will rotate with it.
We therefore modified a wide-field inverted microscope (IX71. Olympus, Japan) using a TIRF objective (Olympus TIRFM 63X/1.45 PlanApo Oil) to allow rotating the plane of incidence [Fig. 2(c)] by changing the position in which the beam is focused in the BFP. 
Two lasers were used: one near the gold particle plasmon resonance (532 nm, Compass C315M. Coherent Inc., USA) and another shifted towards shorter wavelengths (473 nm, VA-I-N-473. Viasho Technology, China). 
The power of the lasers after the objective was ~ 13 $\mu$W.
Circularly polarized light was achieved at the BFP by inserting a quarter and a half wave plate in the beam path adjusted to precompensate for the polarization dependent transmission of the beam splitter, filters and mirrors.
The beam was expanded and filtered to achieve a diffraction limited spot in the BFP. In order to displace the beam in the BFP and therefore change the plane of incidence, a pair of computer controlled galvanometer scanners (SC2000 controller, Minisax amplifier and M2 galvanometer. GSI Group, USA) were used.
The polarization distortion due to the change in the angle of incidence onto the mirrors of the scanners (while moving) was verified to be negligible.
Images of the sample were acquired using a cooled monochrome CCD camera (Alta U32. Apogee Instruments, USA. $2148 \times 1472$ pixels each $6.8 \times 6.8$ $\mu$m$^2$) through a dichroic filter for the fluorescence sample (XF2009 550DCLP. Chroma Technology, USA) or a 30/70 beam splitter for the gold nanoparticles (21009. Chroma Technology, USA).

\subsection{Intrinsic anisotropy determination}
To assure a constant ratio between the two polarizations of the beam and a uniform intensity, the beam needed to be moved on the BFP in a circle centered in the optical axis. Failure to do this would have reduced the dynamic range of the system by introducing an intrinsic anisotropy. To minimize this value, the path of the beam was iteratively modified while measuring the anisotropy (see below) of a diluted solution of Rhodamine 101. The emission of such a sample is independent of the excitation polarization and thus the measured anisotropy can be assigned only to the system. After optimization, the obtained anisotropy for the 532 and 473 channels was 0.06 and 0.05 in a region of $50 \times 50$ $\mu$m$^2$ ($440 \times 440$ pixel$^2$). It is worth noting than these values are five times smaller than the expected anisotropy for a 20 nm homodimer.

\subsection{Image acquisition and processing}
The acquisition process consisted in sequentially imaging at 473 nm and 532 nm while changing the angle $\beta$ in 20 discrete steps over $2\pi$ to sample different polarizations.
An image with both lasers off was also acquired to account for ambient light and dark counts of the camera.
Each image was background corrected and normalized by the excitation power and detection efficiency at the corresponding wavelength.
Mean images ($m_{473}$ and $m_{532}$) were obtained by averaging over all polarizations and, from these, the ratio image $m = m_{532}/m_{473}$ was calculated.

The scattering image at the resonance peak ($m_{532}$) was segmented by Otsu's thresholding and masked with the ratio image thresholded above 1.4 to detect pixels containing gold.
A connected region analysis was performed to keep only those regions with area between 3 and 15 pixels. The upper bound was chosen to be slightly bigger than the airy diffraction limited spot for the system, but still much smaller than the mean distance between gold nanoparticles.
Regions containing single gold nanoparticles should have a constant intensity over the stack of frames acquired for different polarization orientation, while dimers should provide an oscillating signal with period $\pi$.
Therefore, a Fourier analysis [Eq. (\ref{eq:FFT})] was performed. For each pixel, the following coefficients were calculated:

\begin{subequations} \label{eq:FFT}
\begin{align}
\tilde{c}_2 &= \frac{2}{\sum_{\beta} I(\beta) } \sum_{\beta} I(\beta) cos(2\beta) \\
\tilde{s}_2 &= \frac{2}{\sum_{\beta} I(\beta) } \sum_{\beta} I(\beta) sin(2\beta) \\
\eta &= \sqrt{\tilde{c}_n^2 +\tilde{s}_n^2} \\
tan(\delta) &= \frac{\tilde{s}_2}{\tilde{c}_2}
\end{align}
\end{subequations} 
This was done for each wavelength obtaining values for the anisotropy ($\eta_{473}$ and $\eta_{532}$) and orientation ($\delta_{473}$ and $\delta_{532}$) in each pixel. The acquisition and analysis process were repeated for 30 and 10 fields of view of the sample and negative control sample respectively.
Retrieval uncertainty was estimated by performing the same numerical analysis on simulated data calculated by adding two terms to the theoretical response for different dimers.
The first, an oscillating term with $2\pi$ periodicity, emulated a small misalignment that produced a non-constant illumination while rotating the beam in the BFP. 
The amplitude of this term was obtained from the Rhodamine calibration.
The second term simulated coherent background and its values for each pixel were drawn from a Gaussian distribution obtained from the control images.

\figC

\section{Results and discussion}
The need for a two color approach is evident when comparing single wavelength with ratio images: while regions with and without  nanoparticles [Fig. 3(a), circles and squares respectively] were bright due to the high background in the single channel image, only gold nanoparticles were above the threshold in the ratio image.
Importantly, in the negative control stack, no pixel was found above the {\it a priori} defined threshold.
For the 35 regions identified as gold monomers/dimer, a strong correlation between anisotropy and mean value was found as expected [Fig. 3(b)].
The variance over each region for all values was below the corresponding retrieval uncertainty.
A plateau was observed for the 473 nm channel due to the intrinsic anisotropy of the system.
In this set of candidates for dimers, eight points presented an unexpected high individual anisotropy and hence were rejected.
Although the exact origin of this eight scattering centers could not be established, it is worthwhile noting that it is extremely relevant in a tracking experiment to avoid false positives that would severely distort the retrieved information, and this ability to reject scattering centers based on their response is an additional advantage of the technique.
\figD

The recovered scattering parameters were compared with the expected results for a homodimer configuration (Fig. 4) obtaining a good correspondence with the nominal size of the nanoparticles used (20 nm).
Indeed, the mesh shown in Fig. 4 was calculated using only the photophysical and geometrical properties of the dimer (no fitted parameters).
The ability of the technique to blindly recover the correct size of the particles was a cross-check for its reliability. 

The actual configuration of each dimer was obtained by fitting the theoretical model to the experimental values.
The in-plane orientation was directly obtained as a weighted average of these values.
To fit the radii of each particle and the distance between them, the values of $\eta_{473}$, $\eta_{532}$ and $m$ were used.
The values were first fitted using the analytical solution of a homodimer configuration in the dipole-dipole approach, in which the induced dipole moment $\vec{p}$ of each particle in an incident field $\vec{E}_{inc}$ can be expressed as:
\begin{subequations} \label{eq:dipole}
\begin{align}
\vec{p}_{\parallel} = \frac{1}{1-\frac{\alpha}{2d^3\pi}}\epsilon_m\alpha \vec{E}_{inc} \\
\vec{p}_{\perp} = \frac{1}{2+\frac{\alpha}{2d^3\pi}}\epsilon_m\alpha \vec{E}_{inc}
\end{align}
\end{subequations} 
$\epsilon_m$ being the dielectric constant of the medium and $\alpha$ the polarizability of the sphere which is proportional to the cube of its radius \cite{grecco_distance_2006, bohren_absorption_1983}.
This homodimer configuration was used as an initial value in the time consuming iterative process of finding a heterodimer configuration compatible with the experimental data using GMMie calculation.
The traveling wave approximation of the evanescent field was used as the particles are small compared to the decay length of the field \cite{quinten_scattering_1999, wannemacher_resonant_1999} and the collection efficiency is much smaller for the dipole induced in the optical axis orientation than for the one perpendicular to it.

In this way, the two radii, orientation and distance for each dimer were obtained (Fig. 5). 
The distribution of radii was found to be centered in 20 nm, compatible with the nominal size of the particles.
For the interparticle distances, the distribution showed an increase as expected but then a decrease for distances at which the anisotropy is close to the intrinsic anisotropy of the system.  
This mismatch at larger distances is due to the conservative criterion to separate dimers from monomers that fails to identify correctly nanoparticles that couple weakly.
The dimer orientation was uniformly distributed between $-\pi$ and $\pi$, as expected. 
To further test this, we compared the experimental and simulated distributions using a Kolmogorov-Smirnov \cite{massey_kolmogorov-smirnov_1951} statistical test.
The level of significance set at the usual value of 5\%.
The expected distributions (Fig. 5, right column) were obtained from the nominal radius of the nanoparticles and Monte Carlo simulation of the adsorption process. 
The experimental and simulated distributions for radii and orientation were found in close agreement.
For the interparticle distance, the distributions were found compatible when compared up to 70 nm.

\figE

\section{Conclusions}
We have experimentally shown that distance between two nanoparticles, as well as their individual radii, can be obtained by measuring the intensity of spot as a function of the incident polarization.
Additionally, the in-plane orientation of the dimer was obtained with less than 10$^{\circ}$ uncertainty.
The presented method strongly exploits the particular spectroscopic properties of metallic nanoparticles to sense their environment.
It is worth noting that the distance in which the technique is sensitive scales with the radii of the used particles.
By using nanoparticles with radius between 4 and 20 nm, the gap between FRET and standard super-resolution techniques (10 nm to 50 nm) could be bridged.
This fact, together with the ability to recover orientation, makes this approach unique.

A non-uniform anisotropic illumination is the major source of uncertainty as it will mask the anisotropy of the dimers, specially for those in which the distance is much larger than the radius of the particles. This should be properly controlled by measuring an isotropic sample as it was done in this work. Additionally, it is important to mention that various factors such as a non-monodisperse or non-spherical population of particles will have an incidence in the recovery of dimer distance, size and orientation from model based fittings. However, having a multiparametric readout (i.e. $m_{532}$, $m_{473}$, $\eta_{473}$ and $\eta_{532}$) with a non-trivial dependence of the physical parameters (i.e. distance, size, shape) provides a way to control for this and exclude points that do no match the expected relations between photophysical properties.
Such conservative criterion would be recommended for tracking experiments where false negatives have minor impact as they only reduce the amount of information gathered per frame.
If the yield of dimers can be raised and several dozens of dimers can be imaged in the same field of view, we expect that the presented technique will be useful to add information about the relative movement of the two particles to already existing tracking assays.
Numerical simulations showed that if coherent background can be diminished, an order of magnitude (i.e. by the use of broader band excitation source), distance and orientation could be tracked at 100 Hz.
If just the rotation and the movement of the center of mass is desired, the retrieval can be performed much faster as only one wavelength (532 nm) would be necessary after an initial identification of the dimers is made by the two color method.

As other scattering based techniques, the lack of photobleaching constitutes a major advantage of this approach. 
Moreover, the absence of saturation in the light scattering of metallic nanoparticles, as compared to the absorption of fluorescent molecules, provides an acquisition rate only limited by the detector speed.
The combination of these two aspects means that scattering based microscopy does not need to make compromises between experiment length and temporal resolution.

We have also demonstrated that the use of two color imaging can provide an efficient way to detect scattering centers that have plasmon resonances.
Recent work by Olk et al. \cite{olk2008} has shown that upon illumination with a wideband light source, a modulation of the spectra due to far-field effects can be observed as a function of the incident polarization. The combination of the two techniques could lead to a more robust detection of both, orientation and distance.

Finally, the novel illumination setup introduced in this work provides a robust way to change the polarization in TIR,  allowing the implementation of anisotropy based techniques in fluorescence and scattering microscopy. Additionally, the same scheme permits a fast switching between TIR and standard wide-field as well as sweeping of multiple evanescent field penetration depths.

\begin{acknowledgements}
HEG was funded by the Universidad de Buenos Aires.
\end{acknowledgements}



\end{document}